\documentclass[twoside]{article}

\usepackage{lipsum} 
\usepackage{amsmath} 
\usepackage{graphicx} 
\usepackage{multirow} 
\usepackage{fixltx2e}
\usepackage{hyperref}

\usepackage[sc]{mathpazo} 
\usepackage[T1]{fontenc} 
\usepackage{microtype} 

\usepackage[hmarginratio=1:1,top=32mm,columnsep=20pt]{geometry} 
\usepackage{multicol} 
\usepackage[hang, small,labelfont=bf,up,textfont=it,up]{caption} 
\usepackage{booktabs} 
\usepackage{float} 
\usepackage{hyperref} 

\usepackage{lettrine} 
\usepackage{paralist} 

\usepackage{abstract} 

\usepackage{titlesec} 
\titleformat{\section}[block]{\large\scshape\centering\bfseries}{\thesection.}{1em}{} 
\titleformat{\subsection}[block]{\large\bfseries}{\thesubsection.}{1em}{} 
\usepackage{fancyhdr} 
\usepackage{listings}

\makeatletter
\def\lst@PlaceNumber{\ifnum\value{lstnumber}=0\else
  \llap{\normalfont\lst@numberstyle{\thelstnumber}\kern\lst@numbersep}\fi}
\makeatother
\newcommand*\DNumber{\addtocounter{lstnumber}{-1}}

\renewcommand\lstlistingname{Algorithm}

\title{\vspace{-15mm}\fontsize{24pt}{10pt}\selectfont\textbf{Automatic Synthesis of Test Cases to Identify Software Redundancy}} 

\author{
\large
\textsc{Matteo Brunetto}\\[2mm] 
\normalsize University of Milano-Bicocca \\ 
\vspace{-5mm}
}
\date{}

\begin{document}

\maketitle 

\thispagestyle{fancy} 


\begin{abstract}

\noindent Software system can include redundant implementation elements, such as, different
methods that can produce indistinguishable results. This type of redundancy is called `intrinsic' if it is already available in the software, although not intentionally planned. Redundancy can be a key element to increase the reliability of a system. Some fault tolerance and self-healing techniques exploit the redundancy to avoid failures at runtime. Unfortunately, inferring which operations are equivalent manually can be expensive and error prone.

A technique proposed in previous work allows to automatically synthesizes method sequences that are equivalent to a target method. However this technique needs an execution scenario to work. Currently, this execution scenario is generated manually that is expensive and makes the technique hard to use.

This paper proposes a technique to generate execution scenarios for a target method for which we are searching equivalent sequences. The experimental results obtained on the Java class \texttt{Stack} show that the proposed approach correctly generates execution scenarios within reasonable execution time. Besides, the execution scenarios generated allow to maximize the effectiveness of the technique described above.

\end{abstract}


\begin{multicols}{2} 

\section{INTRODUCTION}

\lettrine[nindent=0em,lines=3]{M} odern software is redundant. For example, it can be redundant because redundant components are introduced in the system to obtain fault-tolerant systems. A new aspect is that software systems can be \textit{intrinsically} redundant due to the presence, not deliberately planned, of equivalent code fragments, for example methods or method sequences. Two methods are \textit{equivalent} if they can produce indistinguishable results.
For example, method \texttt{pop()} of the Java class \texttt{Stack} is equivalent to the method sequence \texttt{remove(size()-1)}. Removing the element on the top of the stack leads to the same result as removing the element in the last position.

Equivalent method sequences find many useful applications, from the automatic generation of test oracle~\cite{oracle}, to the design of self-healing system~\cite{workarounds,Gorla2010}.
In these applications, the equivalence must be identified manually. This is a non-trivial activity that may represent an obstacle to the practical applicability of these technique.

Gorla et al. proposed a search-based technique that, given a target method and a initial set of execution scenarios, automatically synthesizes method sequences that are equivalent to the target method.~\cite{sbes}.
This technique has been implemented in a Java prototype tool called SBES that needs an execution scenario to work. The execution scenario is generated manually that is expensive and makes the technique of Gorla et al. hard to use

In this paper we propose a technique that can automate this activity. Given a target method and a target class, the technique automatically generates execution scenarios for the target method.
The generation proceeds in two phases. In the first phase, the goal is to identify a list of methods of the target class that can allow to maximize the effectiveness of SBES. In the second phase, the goal is to generate test cases that contain only methods defined in the first phase and then convert them in execution scenarios.

The technique that we propose in this paper is fully automatic and requires as input only the method for which we are searching equivalent sequences and its class.
Our experiments indicate that the technique is effective in generating execution scenarios, and the execution time is reasonably efficient. On 15 methods of the Java class \textbf{Stack}, our approach always generates at least 1 execution scenario and an average of 26 execution scenarios for each method. Besides, on 6 methods of the Java class \textbf{Stack}, the effectiveness of SBES is similar both using execution scenarios generated automatically with the proposed approach and using execution scenarios
generated manually by experts.

This paper is organized as follows. Section 2 introduces the software redundancy, describes the SBES approach, and discusses the aspects that affect the effectiveness of SBES. Section 3 presents the details of our technique. Section 4 discusses the experimental results obtained to validate the proposed approach. Section 5 summarizes the results presented in the paper and illustrates future developments.


\section{IDENTIFYNG SOFTWARE REDUNDANCY}

\subsection{Software Redundancy}

A software system is \textit{redundant} if the execution of different methods or combinations of methods leads to indistinguishable result. Two executions lead to \textit{indistinguishable} results if they produce the same output and state. For our purpose, we are considering a type of \textit{observational} equivalence where the state produced by the executions may be internally different but not externally distinguishable by analysing the system through its public interface~\cite{Hennessy1980}.
For example, the methods \texttt{get} and \texttt{elementAt} of the Java class \texttt{Stack} return the same object of the stack. We say that two methods are \textit{equivalent} if they produce indistinguishable results for all possible inputs, as in the previous example.

Redundancy can be explicitly added to a software system to increase reliability. There is another type of redundancy that already exists in the software. This type of redundancy is called \textit{intrinsic redundancy} and it is present due to modern design practices like backward compatibility and design for reusability~\cite{Gorla2010}. A library might contain different version of the same component to ensure compatibility with previous system. For example, the Java 7 standard library contains at least 365 methods that are deprecated and that overlap with functionality of newer methods.
Modern development practices induce developers to use external libraries that already implement the needed functionality. It is possible to find several libraries that provide similar functionalities. For example, the Guava library implements collections that are similar to the Java standard library.
In this work we focus on equivalent methods and combinations of methods in the same component.

This type of redundancy finds many interesting applications that span from fault tolerance~\cite{oracle} to self-healing~\cite{workarounds,Gorla2010}. However these techniques rely on manual identification of the equivalence, and this limits their applicability.

Gorla et al. proposed a technique to automatically identify equivalent methods and combinations of methods by exploiting genetic algorithms~\cite{sbes}.

\subsection{Synthesis of Equivalent Methods Sequences}

In this section we describe the technique of Gorla et al.~\cite{sbes} in details. This technique synthesizes a sequence of method invocations that is equivalent to a target method $m$ on a finite set of \textit{execution scenarios} by means of a two-phase iterative process.

It starts with a initial set of execution scenarios that represent a sample of the input space of $m$. The initial set of execution scenarios may be as simple as a single test case.

In the context of Java programs, an execution scenario is a sequence of method invocations that generates objects by means of constructors, operates on such objects by means of public methods, and terminates with an invocation of method $m$. Test Case~\ref{es} is an example of execution scenario for the method \texttt{pop()} of the class \texttt{Stack}.
{\footnotesize
	\begin{align}\label{es}
		<\text{Stack s=new Stack();s.push(1);int result=s.pop()}>
	\end{align}
}%

In the first phase, the technique uses genetic algorithms to generate an equivalent candidate $eq$ for the given set of execution scenarios. In the second phase, the technique validates $eq$ by using genetic algorithms to find a counterexample, which corresponds to an execution sequence for which $eq$ and $m$ are not equivalent. This phase is necessary because the candidate might be valid only on a specific scenario.
If it finds a counterexample, it adds the counterexample to the set of execution scenarios, and it iterates through the first phase looking for a new candidate. Otherwise, if no counterexample is identified, it has successfully synthesizes a method sequence $eq$ that is equivalent to $m$.

This technique has been implemented in a Java prototype tool called SBES.

The experiments indicate that the technique is effective in synthesizing equivalent method sequences within reasonable execution time. On 47 methods belonging in 7 different classes for which equivalent method sequences were know a priori, the technique synthesizes 87\% of the equivalences and one or more equivalent sequences for each target method, with few false positive.~\cite{sbes}

SBES needs an execution scenario to work. Currently, the execution scenario is manually generated
 that is expensive in term of time and makes intrinsic redundancy hard to use.

\subsection{Effectiveness of SBES}\label{method_class}
The effectiveness of SBES depends on the execution scenarios, especially on the size of the \textit{object graphs} in the execution scenarios. The \textit{size} of the object graph is the number of nodes of the object graph whose root is an instance of the target class for which we are searching equivalent sequences.

We experimented with 2 methods of the class \texttt{Stack} as reported in Table~\ref{tab:eff_study}. We ran the experiments by feeding SBES with the class \texttt{Stack}, the target method and 12 execution scenarios with different number of element from 0 to 11. We repeated the experiments 30 times because of the random nature of genetic algorithms.

For each execution scenario, the table shows the following information: \textit{(i)} the average amount of equivalent sequences identified in the 30 runs (column \textbf{Avg}), \textit{(ii)} the precision (\textbf{Prec}) and \textit{(iii)} the number of iterations needed to identify the equivalent sequence \texttt{removeAllElements()} for the method \texttt{clear()} and \texttt{add(Object)} for the method \texttt{push(Object)} (column \textbf{Iterations}).

Precision is defined as the ratio between the number of equivalent sequences correctly synthesized with the approach (true positives) and the total number of sequences deemed as equivalent, which include both the equivalent ones (true positives) and the non-equivalent ones erroneously identified as equivalent by the approach (false positives).
\[
	\frac{\text{true positives}}{\text{true positives} + \text{false positives}}
\]	
\begin{table*}
	\centering
	\caption{Effectiveness of SBES.}
	\label{tab:eff_study}
	\begin{tabular}{|c|c|c|c|}
	\hline
	\multicolumn{4}{|c|}{\textbf{Clear}} \\
	\hline
	\multirow{2}{*}{\textbf{Num. Elements}} & \multicolumn{2}{l|}{\textbf{Equivalence Synthesized}} & 		\multirow{2}{*}{\textbf{Iterations}} \\ \cline{2-3}
                                        & \textbf{Avg}              & \textbf{Prec}              &                                           \\ \hline
	0 & 0 & - & - \\
	\hline
	1 & 2.3 & 0.71 & 2 \\
	\hline
	2 & 2.63 & 0.72 & 2 \\
	\hline
	3 & 2.67 & 0.74 & 1 \\
	\hline
	4 & 2.96 & 0.90 & 1 \\
	\hline
	5 & 3 & 1 & 1 \\
	\hline
	6 & 3 & 1 & 1 \\
	\hline
	7 & 3 & 1 & 1 \\
	\hline
	8 & 3 & 1 & 1 \\
	\hline
	9 & 2.60 & 1 & 1 \\
	\hline
	10 & 2.30 & 1 & 1 \\
	\hline
	11 & 2.13 & 1 & 1 \\
	\hline
	\hline
	\multicolumn{4}{|c|}{\textbf{Push}} \\
	\hline
	\multirow{2}{*}{\textbf{Num. Elements}} & \multicolumn{2}{l|}{\textbf{Equivalence Synthesized}} & 		\multirow{2}{*}{\textbf{Iterations}} \\ \cline{2-3}
                                        & \textbf{Avg}              & \textbf{Prec}              &                                           \\ \hline
	0 & 0 & - & - \\
	\hline
	1 & 0.8 & 0.51 & 2 \\
	\hline
	2 & 0.83 & 0.58 & 2 \\
	\hline
	3 & 1.2 & 0.64 & 2 \\
	\hline
	4 & 1.97 & 0.87 & 1 \\
	\hline
	5 & 2 & 0.90 & 1 \\
	\hline
	6 & 2 & 0.90 & 1 \\
	\hline
	7 & 2 & 0.90 & 1 \\
	\hline
	8 & 2 & 0.90 & 1 \\
	\hline
	9 & 1.7 & 0.90 & 1 \\
	\hline
	10 & 1.6 & 0.95 & 1 \\
	\hline
	11 & 1.23 & 1 & 1 \\
	\hline
	\end{tabular}
\end{table*}

Table~\ref{tab:eff_study} indicates that we need a stack object that contains between 5 and 8 elements to maximize the effectiveness of SBES. Between 5 and 8 elements we obtain the maximum number of equivalent method sequences, with a precision of 100\% and only 1 iteration required.


\section{SYNTEHSIS OF EXECUTION SCENARIOS}

We propose a technique that can automate the creation of initial execution scenarios by means of a two-phase process. We start with the method $m$ for which we are searching equivalent sequences and its class $c$. 
In the first phase, we generate a list of methods of $c$ that can maximizes the effectiveness of SBES (see section~\ref{method_class}).
In the second phase, we generate test cases that contain only methods defined before. Then, these test cases are converted in execution scenarios.

The process for generate execution scenarios for a target method $m$ and a target class $c$ is detailed in Algorithm~\ref{lst:esg_algo}. 

In the first phase (lines 1-8) the algorithm extracts methods of $c$ (line 1) and puts them in a list called \textit{methods}. Then, it searches pure methods (line 2) and methods that decrease the size of the object graph (lines 4-8) and removes them from \textit{methods}. The function \texttt{REMOVE-METHOD} compares the method we want to remove from \textit{methods} with $m$ and if they are equal does not remove it.

Methods can be divided into two categories: \textit{pure methods} and \textit{impure methods}.
A method is pure if it does not change the state of the object. Otherwise, a method is impure if it changes the state of the object~\cite{purity}. 
Impure methods are divided into three categories defined by us: \textit{methods that increase the size of the object graph}, \textit{methods that decrease the size of the object graph} and \textit{methods that change the nodes in the object graph}. 

To maximise the effectiveness of SBES we must exclude pure methods and methods that decrease the size of the object graph from the generation of execution scenarios. Pure methods do not change the size of object graph while methods that decrease the size of the object graph can generate a stack object with less than 5 elements.

In the second phase (lines 9-21) the algorithm generates test cases that contain only the methods defined in \textit{methods} (line 9). Then, these test cases are converted in execution scenarios (line 11). The function \texttt{NORMALIZE-TEST-CASE} reduces the test case by eliminating the unnecessary instructions.

The main iteration terminates when the algorithm generates a set of execution scenarios (line 17) or if there are not execution scenarios (line 19).

\begin{figure*}
\includegraphics[width=\textwidth]{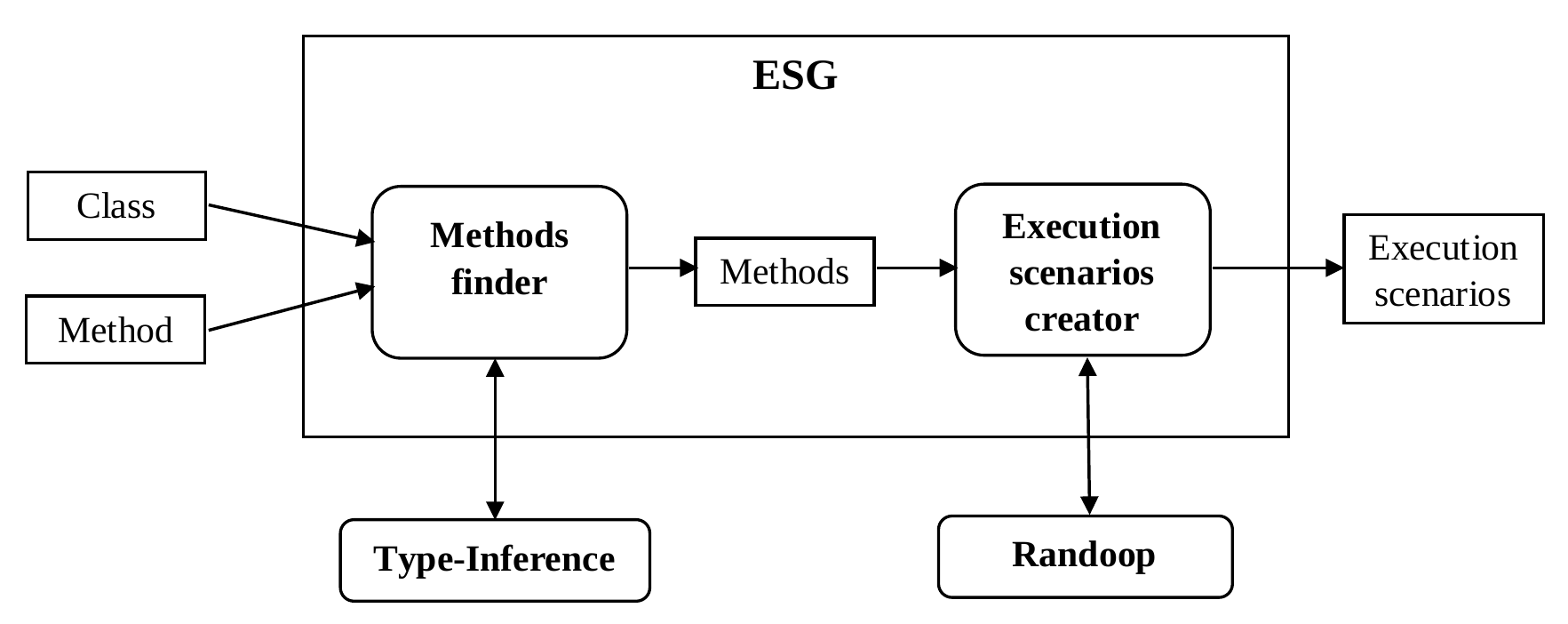}
\caption{Main components of ESG.}\label{FIG:esg_comp}
\end{figure*}

We implemented the algorithm illustrated above in a Java prototype tool called ESG (Execution Scenarios Generator). Figure~\ref{FIG:esg_comp} shows the main components of ESG. The \texttt{Methods Finder} component generates the list of methods by removing pure methods and methods that decrease the size of the object graph. The \texttt{Execution Scenarios Creator} component generates test cases by invoking Randoop and converts them into execution scenarios.
\begin{lstlisting}[caption=%
{Synthesis of execution scenarios.},label=%
lst:esg_algo]
INPUT: c, m
methods := EXTRACT-METHODS(c)
pureMethods := FIND-PURE-METHODS(c)
REMOVE-METHOD(methods,pureMethods)
for each method in methods do
	if DECREMENT-NODE(method) then
		REMOVE-METHOD(methods,method)
	end if
end for
testCases := SYNTHESIZE-TEST-CASES(methods)
for each test in testCases do
	es := NORMALIZE-TEST-CASE(test)
	if NUM-ELEMENTS(es) >= 5 and NUM-ELEMENTS(es) <=8 then
		executionScenarios += es
	end if
end for
if |executionScenarios| != 0 then
	return executionScenarios
else
	return NIL
end if
\end{lstlisting}

In the next sections we detail the key components of ESG, and describe the generation process.

\subsection{First Phase: Generation of the List of Methods}

The first phase generates a list of methods of the target class $c$ that can maximizes the effectiveness of SBES. For this task, the prototype relies on the \texttt{Methods Finder} component. The \texttt{Methods Finder} finds pure methods of $c$ by invoking Type-Inference\footnote{https://code.google.com/p/type-inference/} and removes them from the list of all methods of $c$. Then it finds methods that decrease the size of the object graph by pattern matching approach and removes them from the list of all methods of $c$. For this task, the \texttt{Methods Finder} uses a file manually generated called \textit{blacklist}. This file contains the most common pattern of methods that decrease the size of the object graph. The \texttt{Methods Finder} looks for methods that contain in their name the pattern defined in the \textit{blacklist}.

The following list is an example of pure methods identified for the class \texttt{Stack} and its superclass \texttt{Vector}:
\begin{verbatim}
Stack.empty()
Stack.peek()
Stack.search(Object)
Vector.capacity()
Vector.clone()
Vector.contains(Object)
...
\end{verbatim}

The following list is an example of the common pattern for the class Stack contained in the \textit{blacklist}:
\begin{verbatim}
remove
clear
retain
pop
setSize
\end{verbatim}
The \textit{Methods Finder} uses this list to remove the following methods that decrease the size of the object graph:
\begin{verbatim}
Vector.clear()
Vector.remove(int)
Vector.remove(Object)
Vector.removeAll(Collection)
Vector.removeAllElements()
Vector.removeElement(Object)
Vector.removeElementAt(int)
Vector.retainAll(Collection)
Vector.setSize(int)
Stack.pop()
\end{verbatim}

Given the method \texttt{pop}, the \texttt{Methods Finder} uses the previous lists to generate the list of methods that can maximizes the effectiveness of SBES. The resulting list is reported below:
\begin{verbatim}
Stack.push(Object)
Stack.pop()
Vector.add(int,Object)
Vector.add(Object)
Vector.addAll(int,Collection)
Vector.addAll(Collection)
Vector.addElement(Object)
Vector.set(int,Object)
Vector.insertElementAt(Object,int)
Vector.setElementAt(Object,int)
\end{verbatim}
The reader should notice that method \texttt{pop} is a method that decreases the size of the object graph. However we need this method because it is the method for which we are searching equivalent sequences.

\subsection{Second Phase: Generation of Execution Scenarios}

The second phase generates test cases that contain only the methods defined previously in the first phase and converts them in execution scenarios. For this task, the prototype relies on the \texttt{Execution Scenarios Creator} component. The \texttt{Execution Scenarios Creator} creates test cases from $c$ by invoking Randoop.\footnote{https://code.google.com/p/randoop/} Randoop is an automatic test case generator for Java. Randoop accepts as input a list of the methods to be used to seed the test generation process~\cite{randoop}. 

Given the list generated in the first phase, the \texttt{Execution Scenarios Creator} creates test cases (by invoking Randoop) that belong to one of the following categories:
\begin{enumerate}
	\item Test case that does not include the target method $m$.
\end{enumerate}
	Given $m=$\texttt{pop()} we can see in Test Case~\ref{lst:randoop_first} an example of a test case from the first category.
	\renewcommand\lstlistingname{Test Case}
	\renewcommand{\thelstlisting}{2}
	\begin{lstlisting}[caption=%
{Test case form the first category.},label=%
lst:randoop_first,firstnumber=1]
Stack stack0 = new Stack();
stack0.addElement((Object)10);
Object obj0 = stack0.push((Object)1);
Stack stack1 = new Stack();
boolean b0 = stack0.addAll((Collection)stack1);
stack0.add(0, (Object)(-1));
try {
	Object obj1 = stack0.set(10, (Object)10);
	fail(``Expected exception of type java.lang.ArrayIndexOutOfBoundsException'');
} catch (java.lang.ArrayIndexOutOfBoundsException e) {
	// Expected exception.
}
assertTrue(b0 == false);
\end{lstlisting}
\begin{enumerate}
\setcounter{enumi}{1}
	\item Test case that includes the target method $m$ within a try-catch clause.
\end{enumerate}
We can see in Test Case~\ref{lst:randoop_second} an example of a test case from the second category.
\renewcommand{\thelstlisting}{3}
	\begin{lstlisting}[caption=%
{Test case from the second category.},label=%
lst:randoop_second,firstnumber=1]
Stack stack0 = new Stack();
try {
	Object obj0 = stack0.pop();
	fail(``Expected exception of type java.util.EmptyStackException'');
	} catch (java.util.EmptyStackException e) {
		// Expected exception.
}
\end{lstlisting}
\begin{enumerate}
\setcounter{enumi}{2}
	\item Test case that includes the target method $m$ outside any try-catch clause.
\end{enumerate}
We can see in Test Case~\ref{lst:randoop_third} an example of a test case from the third category.
\renewcommand{\thelstlisting}{4}
	\begin{lstlisting}[caption=%
{Test case from the third category.},label=%
lst:randoop_third,firstnumber=1]
Stack stack0 = new Stack();
stack0.addElement((Object)0);
Stack stack1 = new Stack();
stack1.addElement((Object)10);
Object obj0 = stack1.push((Object)1);
Stack stack2 = new Stack();
stack2.addElement((Object)1);
boolean b0 = stack1.addAll((Collection)stack2);
stack1.add(0, (Object)(-1));
Object obj1 = stack1.pop();
Object obj2 = stack1.push(100);
try {
	Object obj1 = stack0.pop();
	fail(``Expected exception of type java.util.EmptyStackException'');
	} catch (java.util.EmptyStackException e) {
		// Expected exception.
}|\DNumber|

assertTrue(b0 == false);
assertNull(obj0);
\end{lstlisting}

For our purpose we are interested in the test cases of the third category. The first category is discarded because does not contain $m$ while the second category is discarded because this type of test cases cannot be used due to the current limitations of SBES.
The \texttt{Execution Scenarios Creator} discards test cases from the first two categories and converts the remaining test cases in execution scenarios. To convert test cases in execution scenarios, we perform the following operations:
\begin{enumerate}
	\item \textbf{Elimination of instructions after $m$.} 
	
	\noindent{Given the execution scenarios structure, we need to keep $m$ as the last instruction of the execution scenarios. This operation allows us to obtain execution scenarios with $m$ as last instruction.}
	\item \textbf{Reduction to a single object of $c$.} 
	
	\noindent{A test case can contain multiple object of the class $c$ with a few method invocations. Moreover, as we have already discussed, if we have an object with less than 5 elements, this can lead to bad results. This operation allows us to obtain a single object of $c$ with the right number of element.}
	\item \textbf{Specification of the generic objects as integer type and elimination of unnecessary object casting.} 
	
	\noindent{SBES works better when the classes that rely on generics are instantiated on integers. This operation allows us to create execution scenarios that only work on integers.}
	\item \textbf{Elimination of execution scenarios that contain object of type $c$ with less than 5 or more than 8 elements.} 
	
	\noindent{The results in section~\ref{method_class} show that the maximum effectiveness of SBES is obtained with a number of elements between 5 and 8. This operation removes the execution scenarios that do not respect this constrain.}
	\item \textbf{Elimination of execution scenarios that are syntactically equivalent.} 
	
	\noindent{This operation removes the execution scenarios that are syntactically different.}
\end{enumerate}

\renewcommand\lstlistingname{Execution Scenario}
\renewcommand{\thelstlisting}{1}
\begin{lstlisting}[caption=%
{Execution scenarios obtained from test case.},label=%
lst:execution_scenario,firstnumber=1]
Stack<Integer> stack0 = new Stack<Integer>();
stack0.addElement(0);
stack0.addElement(10);
Integer obj0 = stack0.push(1);
stack0.addElement(1);
stack0.add(0,-1);
Integer obj1 = stack0.pop();
\end{lstlisting}

For example, given the Test Case~\ref{lst:randoop_third}, the operations described above allow us to obtain the Execution Scenario~\ref{lst:execution_scenario}.


\section{EVALUATION}

The evaluation of our work aims to answer the following research questions:

\begin{description}
	\item[RQ1] How effectively can the proposed approach generate execution scenarios?
	\item[RQ2] How efficiently can the proposed approach generate execution scenarios?
	\item[RQ3] How effective is the technique compared to manual generation of execution scenarios?
\end{description}

The research question RQ1 deals with the effectiveness of the approach. The research question RQ2 deals with the efficiency of the approach. The research question RQ3 deals with the effectiveness of SBES using execution scenarios generated automatically with the proposed approach.

To answer RQ1 we calculated the number of execution scenarios generated. To answer RQ2 we measured performance as the \textit{time required to generate the list of methods} that can maximizes the effectiveness of SBES (first phase) and the \textit{time required to generate execution scenarios} (second phase), since this two measures directly affect the overall performance of our approach.

For RQ3, we calculated the effectiveness of SBES using execution scenarios generated automatically with the approach. Then we compared the results obtained against the results obtained using execution scenarios generated manually.

\subsection{Experimental setup}
We experimented with the class \texttt{Stack} taken as a representative for the various containers available in the Java standard library.

For RQ1 and RQ2 we experimented with 15 methods of the class \texttt{Stack} as reported in Table \ref{tab:effectiveness_approach}. We ran the experiments by feeding the prototype with the class \texttt{Stack}, the target method and the \textit{blacklist}. We repeated the experiments 5 times because of the random nature of Randoop.

For RQ3 we experimented with 6 methods of the class \texttt{Stack} as reported in Table \ref{tab:effectiveness_auto}. We ran the experiments by feeding SBES with the class \texttt{Stack}, the target method and 5 execution scenarios, randomly chosen, generated by the prototype. We repeated the experiments 30 times because of the random nature of genetic algorithms. Then we compared the results obtain against the results obtained from the original article about SBES~\cite{sbes}.

\subsection{Results}
In this section we discuss the experimental results.
Table \ref{tab:effectiveness_approach} summarizes the results of the experiment for RQ1.
For each of the analysed methods, the table shows the following information: \textit{(i)} the minimal number of execution scenarios generated with a single run (column \textbf{Min}), \textit{(ii)} the maximum number of execution scenarios generated with a single run (column \textbf{Max}) and \textit{(iii)} the average amount of execution scenarios generated in the 5 runs (column \textbf{Avg}).

Table \ref{tab:effectiveness_approach} shows a very interesting results because the approach always generates at least 1 execution scenario and on average it generates a large number of execution scenarios, about 26 per method.

In summary, we can answer positively to research question RQ1:
\\\\
\framebox{\parbox{\dimexpr\linewidth-2\fboxsep-2\fboxrule}{\itshape%
\textbf{RQ1}: The proposed approach can correctly generate one or more execution scenarios per method.}}
\\\\

Table \ref{tab:efficiency_approach} summarizes the results of the experiment for RQ2. For each method we calculated the median of the results obtained. The results obtained were similar between them so we show in Table \ref{tab:efficiency_approach} the results obtained for the method \texttt{pop}.

For this method, the table shows the following information: \textit{(i)} the time required to identify pure methods (column \textbf{Pure}), \textit{(ii)} the time required to create the list of methods (column \textbf{Methods}), \textit{(iii)} the time required to generate test cases (column \textbf{Test}), \textit{(iv)} the time required to convert test cases in execution scenarios (column \textbf{Convert}) and \textit{(v)} the total time required by the approach to generate execute scenarios (column \textbf{Tot}).
\begin{table}[H]
	\centering
	\caption{Effectiveness of the approach.}
	\label{tab:effectiveness_approach}
	\begin{tabular}{|l|c|c|c|}
	\hline
	\textbf{Method} & \textbf{Min} & \textbf{Max} & \textbf{Avg}\\
	\hline
	add(int,Object) & 8 & 24 & 16.4\\
	\hline
	add(Object) & 6 & 37 & 24\\
	\hline
	addElement(Object) & 4 & 30 & 16.8\\
	\hline
	clear() & 1 & 45 & 32.6\\
	\hline
	elementAt(int) & 14 & 24 & 20\\
	\hline
	firstElement() & 20 & 41 & 28.8\\
	\hline
	get(int) & 20 & 27 & 22.6\\
	\hline
	indexOf(Object) & 12 & 57 & 33.8\\
	\hline
	lastElement() & 14 & 41 & 24.2\\
	\hline
	peek() & 23 & 50 & 36.4\\
	\hline
	pop() & 25 & 56 & 40\\
	\hline
	push(Object) & 11 & 43 & 28.4\\
	\hline
	remove(Object) & 9 & 51 & 31.6\\
	\hline
	remove(int) & 18 & 25 & 22\\
	\hline
	set(int,Object) & 16 & 25 & 19.2\\
	\hline
	\end{tabular}
\end{table}

\begin{table}[H]
\centering
\caption{Efficiency of the approach.}
\label{tab:efficiency_approach}
\begin{tabular}{|c|c|c|c|c|}
\hline
\multicolumn{2}{|c|}{\textbf{I Phase}} & \multicolumn{2}{|c|}{\textbf{II Phase}} & \multirow{2}{*}{\textbf{Tot}} \\ \cline{1-4}
\textbf{Pure}         & \textbf{Methods}        & \textbf{Test}         & \textbf{Convert}         &                         \\ \hline
10.5s        & 0.06s         & 11s          & 10s           & 31.56s                  \\ \hline
\end{tabular}
\end{table}

The execution time is acceptable and less than the time required to manually generate 26 execution scenarios.
Besides the results show that the second phase takes 67\% of the total time.
Hence, we can answer positively to research question RQ2:
\\\\
\framebox{\parbox{\dimexpr\linewidth-2\fboxsep-2\fboxrule}{\itshape%
\textbf{RQ2}: The proposed approach requires a total execution time that is acceptable.}}
\\\\

Table \ref{tab:effectiveness_auto} reports the data about the effectiveness of SBES using execution scenarios generated automatically with the approach. Table \ref{tab:effectiveness_manual} reports the data about the effectiveness of SBES using execution scenarios generated manually by experts.
For each methods, the table shows the following information: \textit{(i)} the number of minimal equivalent sequences identified with manual inspection (column \textbf{Tot}), which we use
as baseline, \textit{(ii)} the amount of equivalent sequences automatically synthesized in at least one run (column\textbf{Max\textsubscript{t}}), \textit{(iii)} the maximum amount of equivalent sequences synthesized with a single run (column \textbf{Max\textsubscript{r}}), \textit{(iv)} the average amount of equivalent sequences identified in the 30 runs (column \textbf{Avg}), \textit{(v)} the precision (\textbf{Prec}) and the recall (\textbf{Rec}) computed over the 30 runs~\cite{sbes}.

Recall is defined as the ratio between the number of equivalent sequences correctly synthesized with the approach (true positives) and the total number of equivalent sequences, which include both the ones correctly synthesized (true positives) and the ones that the approach fails to synthesize
(false negatives). 
\[
	\frac{\text{true positives}}{\text{true positives} + \text{false negatives}}
\]

Precision is defined as the ratio between the number of equivalent sequences correctly synthesized with the approach (true positives) and the total number of sequences deemed as equivalent, which include both the equivalent ones (true positives) and the non-equivalent ones erroneously identified as equivalent by the approach (false positives).
\[
	\frac{\text{true positives}}{\text{true positives} + \text{false positives}}
\]	

\begin{table*}
	\centering
	\caption{Effectiveness of the automatic approach.}
	\label{tab:effectiveness_auto}
	\begin{tabular}{|l|c|c|c|c|c|c|}
	\hline
	\textbf{Method} & \textbf{Tot} & \textbf{Max\textsubscript{t}} & \textbf{Max\textsubscript{r}} & \textbf{Avg} & \textbf{Prec} & \textbf{Rec}\\
	\hline
	addElement & 6 & 4 & 2.8 & 2.04 & 0.99 & 0.70 \\
	\hline
	clear & 3 & 3 & 3 & 2.73 & 0.91 & 1 \\
	\hline
	firstElement & 2 & 2 & 2 & 1.40 & 0.77 & 1 \\
	\hline
	peek & 2 & 2 & 2 & 1 & 0.77 & 1 \\
	\hline
	push & 6 & 2 & 2 & 2 & 0.94 & 0.33 \\
	\hline
	remove(Object) & 4 & 1.8 & 1 & 0.68 & 0.86 & 0.45 \\
	\hline
	\end{tabular}
\end{table*}
\begin{table*}
	\centering
	\caption{Effectiveness of the manual approach.}
	\label{tab:effectiveness_manual}
		\begin{tabular}{|l|c|c|c|c|c|c|}
	\hline
	\textbf{Method} & \textbf{Tot} & \textbf{Max\textsubscript{t}} & \textbf{Max\textsubscript{r}} & \textbf{Avg} & \textbf{Prec} & \textbf{Rec}\\
	\hline
	addElement & 6 & 4 & 3 & 2.17 & 1 & 0.70 \\
	\hline
	clear & 3 & 3 & 3 & 2.77 & 0.99 & 1 \\
	\hline
	firstElement & 2 & 2 & 2 & 1.57 & 0.89 & 1 \\
	\hline
	peek & 2 & 2 & 2 & 1.23 & 0.97 & 1 \\
	\hline
	push & 6 & 2 & 2 & 2 & 1 & 0.33 \\
	\hline
	remove(Object) & 4 & 2 & 1 & 0.80 & 0.92 & 0.50 \\
	\hline
	\end{tabular}
\end{table*}

The results show that for the values \textbf{Max\textsubscript{t}}, \textbf{Max\textsubscript{r}} and \textbf{Rec} our approach is slightly lower than the manual approach. Besides, for the values \textbf{Prec} and \textbf{Avg} our approach is lower than the manual approach of 9\% for the \textbf{Prec} values and 11\% for the \textbf{Avg} values.

We have examined execution scenarios from the two approaches to discover the cause of the discordant results. 
We have discovered another aspect that increments the effectiveness of SBES. This aspect is the \textit{heterogeneity} of the values of the stack. The more the values are heterogeneous, the greater the effectiveness of SBES.

We have repeated the previous experiments by introducing heterogeneous values in the stack. Table~\ref{tab:effectiveness_auto2} shows the results obtained.
\begin{table*}
	\centering
	\caption{Effectiveness of the automatic approach with the new aspect.}
	\label{tab:effectiveness_auto2}
		\begin{tabular}{|l|c|c|c|c|c|c|}
	\hline
	\textbf{Method} & \textbf{Tot} & \textbf{Max\textsubscript{t}} & \textbf{Max\textsubscript{r}} & \textbf{Avg} & \textbf{Prec} & \textbf{Rec}\\
	\hline
	addElement & 6 & 4 & 3 & 2.15 & 0.99 & 0.70 \\
	\hline
	clear & 3 & 3 & 3 & 2.73 & 0.99 & 1 \\
	\hline
	firstElement & 2 & 2 & 2 & 1.54 & 0.85 & 1 \\
	\hline
	peek & 2 & 2 & 2 & 1.20 & 0.97 & 1 \\
	\hline
	push & 6 & 3 & 2.4 & 2 & 0.97 & 0.50 \\
	\hline
	remove(Object) & 4 & 2 & 1 & 0.80 & 0.92 & 0.50 \\
	\hline
	\end{tabular}
\end{table*}

The results show that now for the values \textbf{Max\textsubscript{t}}, \textbf{Max\textsubscript{r}} and \textbf{Rec} our approach is equal than the manual approach and for the method \texttt{push} we get even better results. Besides, for the values \textbf{Prec} and \textbf{Avg} our approach is lower than the manual approach but only of 1,33\% for the \textbf{Prec} values and 2\% for the \textbf{Avg} values.

In summary, we can answer positively to research question RQ3:
\\\\
\framebox{\parbox{\dimexpr\linewidth-2\fboxsep-2\fboxrule}{\itshape%
\textbf{RQ3}: The effectiveness of SBES is similar both using execution scenarios generated automatically with the approach and using execution scenarios generated manually by experts.}}

\section{CONCLUSIONS}
Software redundancy that already exists in the software, called intrinsic redundancy, finds many interesting applications that span from fault tolerance to self-healing. However these techniques rely on manual identification of the equivalence, and this limits their applicability.

Gorla et al. proposed a technique that automatically identifies equivalent methods and combinations of methods by exploiting genetic algorithms, but needs an execution scenario to work. Currently, the execution scenario is manually generated that is expensive and makes the technique hard to use.

In this paper, we presented a novel technique that generates execution scenarios. The technique is fully automatic and applies to any methods. We reported the experimental results obtained with a prototype that implements the approach. The results obtained for the Java class \texttt{Stack} are encouraging. We can automatically generate at least 1 execution scenario and an average of 26 execution scenarios for each method in a total execution time that is acceptable and less than the time required to manually generate 26 execution scenarios. Besides, the effectiveness of SBES is similar both using execution scenarios generated automatically with the approach and using execution scenarios generated manually by experts.

We are currently working on increasing automation with respect to the definition of the \textit{blacklist} file. We are also working on evaluating the approach with new case studies to obtain a more general estimate of the effectiveness and efficiency of the approach.


\bibliographystyle{plain} 
\bibliography{biblio}

\begin{thebibliography}{1}

\bibitem{oracle}
A.~Carzaniga, A.~Goffi, A.~Gorla, A.~Mattavelli, and M.~Pezz\`{e}.
\newblock Cross-checking oracles from intrinsic software redundancy.
\newblock In {\em ICSE'14:Proceedings of the International Conference on
  Software Engineering}, pages 931--942. ACM, 2014.

\bibitem{workarounds}
A.~Carzaniga, A.~Gorla, A.~Mattavelli, N.~Perino, and M.~Pezz\`{e}.
\newblock Automatic recovery from runtime failures.
\newblock In {\em ICSE'13:Proceedings of the International Conference on
  Software Engineering}, pages 782--791. ACM, 2013.

\bibitem{Gorla2010}
A.~Carzaniga, A.~Gorla, N.~Perino, and M.~Pezz\`{e}.
\newblock Automatic workarounds for web applications.
\newblock In {\em FSE'10:Proceedings of the International Symposium on the
  Foundations of Software Engineering}, pages 237--246. ACM, 2010.

\bibitem{sbes}
A.~Gorla, A.~Goffi, A.~Mattavelli, M.~Pezz{\`e}, and P.~Tonella.
\newblock Search-based synthesis of equivalent method sequences.
\newblock In {\em FSE'14:Proceedings of the International Symposium on the
  Foundations of Software Engineering}, pages~--. ACM, 2014.

\bibitem{Hennessy1980}
M.~Hennessy and M.~Milner.
\newblock On observing nondeterminism and concurrency.
\newblock In {\em Proceedings of the 7th Internetional Colloquium on Automata,
  Languages and Programming (ICALP)}, pages 299--309. Springer, 1980.

\bibitem{randoop}
C.~Pacheco, S.~K. Lahiri, M.~D. Ernst, and T.~Ball.
\newblock Feedback-directed random test generation.
\newblock In {\em ICSE'07:Proceedings of the International Conference on
  Software Engineering}, pages 75--84. IEEE Computer Society, 2007.

\bibitem{purity}
A.~S\u{a}lcianu and M.~Rinard.
\newblock Purity and side effect analysis for java programs.
\newblock In {\em VMCAI'05:Proceedings of the International Conference on
  Verification, Model Checking, and Abstract Interpretation}, pages 199--215.
  Springer, 2005.

\end{thebibliography}


\end{multicols}

\end{document}